# Complex Spin Structures and Origin of the Magneto-Electric Coupling in $YMn_2O_5$

J.-H. Kim[1], S.-H. Lee[1*], S. I. Park[2], M. Kenzelmann[3], J. Schefer[3], J.-H. Chung[4], C. F. Majkrzak[5], M. Takeda[6], S. Wakimoto[6], S. Y. Park[7], S-W. Cheong[7], M. Matsuda[6], H. Kimura[8], Y. Noda[8], K. Kakurai[6]

[1] Department of Physics, University of Virginia, Charlottesville, VA 22904-4714, USA

[2] Neutron Science Division, Korea Atomic Energy Research Institute, Daejeon, Korea

[3] Laboratory for Neutron Scattering ETH Zurich and Paul Scherrer Institute, CH-5232 Villigen PSI, Switzerland

[4] Department of Physics, University of Korea, Seoul 136-713, Korea

[5] NIST Center for Neutron Research, National Institute of Standards and Technology, Gaithersburg, MD 20899, USA

[6] Quantum Beam Science Directorate, Japan Atomic Energy Agency, Tokai, Ibaraki 319-1195, Japan

[7] Rutgers Center for Emergent Materials and Department of Physics & Astronomy, Rutgers University, Piscataway, NJ 08854, USA

[8] Institute of Multidisciplinary Research for Advanced Materials, Tohoku University, Sendai 980-8577, Japan

*Abstract:* **By combining neutron four-circle diffraction and polarized neutron diffraction techniques we have determined the complex spin structures of a multiferroic, $YMn_2O_5$, that exhibits two ferroelectric phases at low temperatures. The obtained magnetic structure has spiral components in both the low temperature ferroelectric phases that are magnetically commensurate and incommensurate, respectively. Among proposed microscopic theories for the magneto-electric coupling, our results are consistent with both the spin-current model that requires a magnetic transverse (cycloidal) spiral structure in order to** **induce a spontaneous electric polarization and the magneto-restriction model. Our results also explain why the electric polarization changes at the commensurate-to-incommensurate phase transition.**

Magnetic multiferroic materials become simultaneously ferroelectric and magnetic at low temperatures and are thus attractive for use in technological devices that can exploit both sets of properties[1-5]. They usually undergo multiple magnetic and electronic phase transitions upon cooling. For instance, when temperature is lowered, $AMn_2O_5$ (A = Tb, Y, Er) [2,16-20] first changes from a paramagnetic and paraelectric (HTPM-PE) to a magnetically incommensurate and paraelectric (HTI) phase. Upon further cooling, a commensurate magnetic ordering appears with a characteristic wave vector, $k_C$ = (1/2,0,1/4) and a spontaneous electric polarization, ***P***, is observed along the +b-axis, entering the intermediate temperature commensurate-ferroelectric (ITC-FE) phase. At an even lower temperature, the ITC-FE state is replaced by a low temperature incommensurate (with $k_{IC}$ = (1/2+$\delta_a$,0,1/4+ $\delta_c$)) and ferroelectric state (LTI-FE) with a much weaker ***P***.

Various theories have been proposed to explain the magneto-electric coupling in magnetic multiferroics[6-12]: the symmetry-based phenomenological Ginzburg-Landau theory[8], spin-current mechanism[6,7,9], magneto-striction mechanism[11], and delocalized spin density wave model[12]. In the spin-current model, the spontaneous electric polarization, ***P***, occurs when the magnetic ground state has a non-collinear transverse (cycloidal) spiral structure and yields a non-zero spin current $e_{ij} \times (S_i \times S_j)$ where $e_{ij} = \frac{r_i - r_j}{|r_i - r_j|}$ is the unit vector connecting the two magnetic ions: $\boldsymbol{P} = \alpha \boldsymbol{e}_{ij} \times (\boldsymbol{S}_i \times \boldsymbol{S}_j)$.

This model explains the magneto-electric phenomena found in many different materials such as TbMnO$_3$ [1,13,14] and CoCr$_2$O$_4$ [15]. On the other hand, in the magneto-striction model $\boldsymbol{P} \propto \boldsymbol{S}_i \cdot \boldsymbol{S}_j$ can occur for a collinear spin structure that has ↑↑↓↓ configuration. Previous powder neutron diffraction studies on AMn$_2$O$_5$ (A = Y and Tb) [16,17] have reported nearly collinear spin structures for their ferroelectric phases and thus presented these systems as where the magneto-striction mechanism not the spin-current model is at work. In the previous study, the decrease in $\boldsymbol{P}$ at the transition from the ITC-FE to the LTI-FE phase was attributed to a magnetic transition from a structure with magnetic moments of similar amplitudes to an amplitude-modulated sinusoidal spin structure. The same result was also referred as experimental evidence for a recent theory based on delocalized spin density waves.[12]

The question is, however, in what way one can uniquely determine a complex spin structure? For instance, there are eight Mn ions in AMn$_2$O$_5$ at two distinct sites in a chemical unit cell: one is the octahedral site occupied by the Mn$^{4+}$ ions and the other is the pyramid site occupied by the Mn$^{3+}$ ions. The neighboring Mn$^{4+}$O$_6$ octahedra along the c-axis share edges and form a chain. The Mn$^{4+}$O$_6$ octahedra share corners with neighboring Mn$^{3+}$O$_5$ trigonal bipyramids and form a zigzag chain in the ab-plane. Thus, determining the spin structure of YMn$_2$O$_5$ requires optimization of 49 parameters and powder neutron diffraction alone is not sufficient to correctly find such a complex structure. Due to its intrinsic powder averaging, powder diffraction data provide limited information and are especially lacking three-dimensional directional information on the wave vector. Four-circle diffraction (FCD) from a single crystal, on the other hand, can probe each wave vector separately and thus provide much more detailed information than powder diffraction. The FCD technique has been instrumental in studying many different magnetic multiferroics. However, we will see later that the FCD is not enough when it comes to complex spin structures as in $AMn_2O_5$.

Here, we demonstrate that combination of four-circle diffraction (FCD) and polarized neutron diffraction (PND) techniques can correctly lead to complex spin structures of $YMn_2O_5$ which turn out to have spin structures in both the ITC-FE and the LTI-FE phase. Our spiral structure of the ITC-FE phase is consistent with recent FCD results[21,22] with some minor modifications: the bc-components of $Mn^{4+}$ ions form a transverse (cycloidal) spiral along the c-axis which, according to the spin-current mechanism, can induce an electric polarization, ***P***, along the +b direction as observed experimentally. We show that the LTI-FE phase also has a magnetic spiral structure that is much more complex than the ITC-FE phase and is very different from the sinusoidal spin structure proposed by the previous powder neutron diffraction study[17]: the ab-components of the $Mn^{4+}$ moments form a longitudinal spiral along the c-axis, the ac- and bc-components form transverse (cycloidal) spirals along the c-axis, and the ab-components of $Mn^{3+}$ and $Mn^{4+}$ moments form cycloidal spirals along the a-axis. When applied to the LTI-FE structure, both the spin-current and the magneto-striction model can explain the decrease of ***P*** in the LTI-FE phase. Other theoretical implications of the new spin structures of $YMn_2O_5$ are also discussed.

A 3g single crystal of $YMn_2O_5$ was used for our neutron four-circle diffraction (FCD) and polarized neutron diffraction (PND) measurements. In the FCD measurements relative spin directions and magnitudes in a magnetic system give arise to relative intensities of magnetic Bragg reflections, while in the PND measurements they give rise to different intensities in the non-spin-flip and the spin-flip channels at each reflection. Thus, in the FCD technique, enough information for the structure determination can only be obtained by measuring a large number of reflections. As the complexity of the structure increases, the required number of reflections grows. In the PND technique, on the other hand, information about particular spin directions and magnitudes can be achieved at each reflection. Thus, the two techniques can be complimentary and powerful, when combined, in determining a complex spin structure. Our FCD measurements were performed at the Paul Scherrer Institute to collect about 300 magnetic reflections in each FE phase. Our PND measurements were done at two neutron facilities using two different experimental configurations: at the National Institute of Standards and Technology, the conventional PND measurements with two sets of transmission neutron polarizers and spin flippers before and after the sample, and a vertical guide field along the beam path to maintain the selected spin state of neutrons. At the Japan Atomic Energy Agency, the three-dimensional polarization analysis (CRYOPAD) technique was used. In both PND measurements, the crystal was aligned in the (h0l) scattering plane. Thus, in the conventional PND measurements, the b-components of the magnetic moments go to the non-spin-flip channel while the ac-components to the spin-flip channel. The polarization efficiency, i.e. the fraction of neutrons that are polarized along the selected directions, was measured at the nuclear reflections to be 82% for the conventional PND and 94% for the CRYOPAD measurements. The experimental data from the two techniques were the same within the experimental error.

Black symbols in Fig. 1 that are buried underneath red symbols are the results of our FCD and PND measurements performed at 25 K (ITC-FE phase) (Fig. 1 (a) and (b)) and 10 K (LT-FE phase) (Fig. 1 (c) and (d)). Firstly, Fig. 1 (a) and (c) clearly show that the nearly-collinear coplanar spin structures proposed by the previous neutron powder diffraction study cannot reproduce our FCD data in both phases (blue symbols in those figures). This clearly illustrates the limitation of the neutron powder diffraction technique in determining complex spin structures and calls for more advanced techniques. Secondly, the previously proposed weakly spiral spin structure based on the FCD technique does not perfectly reproduce our PND data (cyan diamonds in Fig. 1 (b)). This tells us that there might be several spin structures that can reproduce the FCD data yet yield different polarized neutron data. Indeed, when we fitted only our FCD data for the ITC-FE phase, the optimal spin structure obtained was a structure that is similar but not the same as the previously reported weakly spiral spin structure, and did not fit our PND data. In order to obtain a spin structure that reproduces both the FCD and PND data well, we fitted the FCD and PND data *simultaneously*. To our surprise, once the PND data were included, a direct least-square refinement quickly converged after a few cycles to the spin structures that reproduced perfectly both FCD and PND data for the ITC-FE and LTI-FE phases, as can be seen by red symbols in Fig. 1.

Magnetic moments of $Mn^{3+}$ and $Mn^{4+}$ ions of the obtained magnetic structures for the ITC-FE and LTI-FE phases are listed in Tables I and II, respectively, and are illustrated in Fig. 2 and 3. For the ITC-FE phase, the magnetic structure is almost the same as the previously reported structure[21] based on FCD data with some slight modifications: the ab-components of $Mn^{4+}$ moments are almost collinear along the c-axis (Fig. 2 (a)), the ac-components (Fig. 2 (b)) and bc-components (Fig. 2 (c)) of $Mn^{4+}$ ions form cycloidal spirals along the c-axis that can induce spin currents along a- and b-axis, respectively. The zigzag chain formed by $Mn^{4+}$ and $Mn^{3+}$ on the ab-plane is also nearly collinear (Fig. 2 (d)). According to the spin current model, the ac- and bc-spirals can give rise to electric polarization, ***P***. ***P*** induced by the ac-spirals of neighboring chains are, however, in opposite directions and thus cancel with each other (Fig. 2 (b)), while ***P*** induced by the bc-spirals are in the same +b direction (Fig. 2 (c)), resulting in the bulk electric polarization along the +b direction as observed experimentally. On the other hand, the magneto-restriction model can be applied to the chain of $Mn^{4+}$-$Mn^{3+}$-$Mn^{3+}$-$Mn^{4+}$ ions along the b-axis that are more or less parallel and are arranged in a sequence of ++-+, which leads to a strong ***P*** along the b-axis.[17]

For the LTI-FE phase, the only magnetic structure proposed so far is the one that was done by a neutron powder diffraction study[17]. In that model, in each $Mn^{4+}$ chain along the c-axis the almost collinear $Mn^{4+}$ moments modulate sinusoidally. In the ab-plane, $Mn^{4+}$ and $Mn^{3+}$ moments are nearly collinear and coplanar, and along the $Mn^{4+}$ - $Mn^{3+}$ zigzag chains, their magnitudes follow a sinusoidal modulation. Our new incommensurate magnetic structure, however, is quite different. As shown in Fig. 3 (a)-(d), the magnitudes of $Mn^{4+}$ and $Mn^{3+}$ moments hardly change along any axis. Instead, they form complex spiral structures along the a- and c-axes. Firstly, the ab-components of the $Mn^{4+}$ moments rotate about the c-axis (see Fig. 3 (a)), forming a longitudinal spiral. This does not give rise to any electric polarization because it yields zero spin current $e_{ij} \times (S_i \times S_j) = 0$. Secondly, the ac- and bc-components form cycloidal spirals along the c-axis (Fig. 3 (b) and (c)) as in the ITC-FE phase. In the LTI-FE phase, the electric polarizations due to the neighboring $Mn^{4+}$ chains do not cancel either along the a-axis or along the b-direction. Thirdly, the ab-components of the moments form a cycloidal spiral along the a-axis (see Fig. 3 (d)), which induces electric polarization in the ab-plane. When all $P$s from the three types of cycloidal spirals are combined, the bulk $P$ would have a nonvanishing a-component as well as a b-component. As shown in Fig. 3, when we add all $P$s induced by the two chains ((b)-(d)) the b-component of the resulting total $P$ of the LTI-FE phase is much weaker than that of the ITC-FE phase, as experimentally observed. Our results also predict, that when measured along the a-axis, the electric polarization will appear in the ITC-FE phase, even though it will be weak. Application of the magneto-restriction model on our IC spin structure also leads to a weak $P$ along b because the $Mn^{4+}$-$Mn^{3+}$ moments along the b-axis are now almost orthogonal.

Our finding that $YMn_2O_5$ has complex spiral structures in both ITC-FE and LTI-FE phases has important theoretical implications in the field of multiferroics. Firstly, the spiral structures of $YMn_2O_5$ are consistent with both the spin-current model and the magneto-striction model. Secondly, according to the spin-current model, the incommensurate structure of $YMn_2O_5$ may induce weak $P$ along the a-axis as well as along the b-axis. This will have to be checked by measuring $P$ as a function of temperature with an application of an external electric field along the a-axis. According to a symmetry-based Ginzburg-Landau theory[23], when the fourth order terms of the order parameters or Umklapp magnetoelectric interactions are included in the magnetoelectric interaction Hamiltonian for $k_{IC}$ = (1/2+$\delta_a$,0,1/4+ $\delta_c$) with nonzero $\delta_a$, the weak spontaneous polarization $P$ should appear in all directions. Indeed, electric polarization has recently been observed along the a-axis in an incommensurate phase of $TmMn_2O_5$[24]. This might be consistent with our finding that an incommensurate spin structure of $AMn_2O_5$ can induce $P$ along the a-axis. Of course, we need to keep in mind that only two types of chains in this compound were considered for the spin-currents: the $Mn^{4+}$ chains along the c-axis and the $Mn^{4+}$-$Mn^{3+}$ zigzag chains in the ab-plane. It should also be noted that the spin-current model is a simple theory of the magnetoelectric effect of two neighboring magnetic moments and in our estimation of $P = \alpha e_{ij} \times (S_i \times S_j)$ the coefficient $\alpha$ was assumed to be the same for the two different types of chains. The application to a real material would require some complicated modifications to the theory. Nonetheless, our simple analysis explains why $P$ weakens considerably at the ITC-FE to the LTI-FE phase transition as observed experimentally. It is also possible that both the mechanisms might be in play in this complex system. The complex magnetic structures of $YMn_2O_5$ reported here will impose a strict restriction on the theoretical endeavor.

## Acknowledgements

We thank L. C. Chapon, P.G. Radaelli, and A. B. Harris for helpful discussions. Work at Rutgers was supported by the NSF-DMR-0520471.

**Table 1:** Magnetic moments of $Mn^{3+}$ and $Mn^{4+}$ ions in one quarter of the magnetic unit cell of the ITC-FE phase ($k_C = (1/2,0,1/4)$) of $YMn_2O_5$, obtained by fitting the four-circle diffraction (FCD) and the polarized neutron diffraction (PND) data at 25 K. Position is the coordinate of the magnetic ion in the unit of the lattice constants, and $M_a$, $M_b$, $M_c$, and M are the a-, b-, c-components and the magnitude of the moments, respectively. The moments in next unit cells separated by (1,0,0) or (0,0,2) are opposite to the ones listed here, while those in the unit cell separated by (1,0,2) are the same as the listed ones. Errors for all parameters are given within parenthesis.

|  | Position | $M_a$ ($\mu_B$) | $M_b$ ($\mu_B$) | $M_c$ ($\mu_B$) | M ($\mu_B$) |
|---|---|---|---|---|---|
| $Mn^{3+}$ | (0.088, 0.851, 0.5) | 2.1(3) | 0.76(6) | -0.64(5) | 2.3(3) |
| $Mn^{3+}$ | (0.912, 0.149, 0.5) | 1.9(3) | 0.27(6) | -0.78(5) | 2.1(3) |
| $Mn^{3+}$ | (0.412, 0.351, 0.5) | -2.1(3) | 0.68(7) | -0.47(3) | 2.3(3) |
| $Mn^{3+}$ | (0.588, 0.649, 0.5) | 2.1(3) | -0.44(7) | 0.60(3) | 2.2(3) |
| $Mn^{3+}$ | (0.088, 0.851, 1.5) | -3.1(2) | -0.43(6) | -0.22(7) | 3.1(2) |
| $Mn^{3+}$ | (0.912, 0.149, 1.5) | -2.9(2) | -0.65(6) | -0.59(7) | 3.1(2) |
| $Mn^{3+}$ | (0.412, 0.351, 1.5) | 3.3(2) | -0.36(6) | -0.26(6) | 3.3(2) |
| $Mn^{3+}$ | (0.588, 0.649, 1.5) | -2.9(2) | 0.83(6) | 0.26(6) | 3.0(2) |
| $Mn^{4+}$ | (0.5, 0, 0.255) | -1.6(2) | -0.72(4) | 0.41(8) | 1.8(1) |
| $Mn^{4+}$ | (0, 0.5, 0.255) | 1.7(2) | -0.48(2) | 0.11(7) | 1.8(2) |
| $Mn^{4+}$ | (0, 0.5, 0.745) | 1.1(2) | -0.46(6) | 0.67(2) | 1.4(2) |
| $Mn^{4+}$ | (0.5, 0, 0.745) | -1.0(2) | -0.50(6) | 0.89(2) | 1.4(2) |
| $Mn^{4+}$ | (0.5, 0, 1.255) | 1.6(2) | 0.37(8) | 0.75(5) | 1.8(2) |
| $Mn^{4+}$ | (0, 0.5, 1.255) | -1.6(2) | 0.19(5) | 0.68(2) | 1.7(1) |
| $Mn^{4+}$ | (0, 0.5, 1.745) | -2.3(1) | 0.59(5) | -0.11(7) | 2.4(1) |
| $Mn^{4+}$ | (0.5, 0, 1.745) | 2.1(1) | 0.53(6) | 0.17(9) | 2.2(1) |

**Table 2:** Magnetic parameters for the magnetic structure of the LTI-FE phase ($k_{IC}$ = (1/2+$\delta_a$,0,1/4+$\delta_c$) with $\delta_a$ = -0.02 and $\delta_c$ = 0.038) of YMn$_2$O$_5$, obtained by fitting the four-circle diffraction (FCD) and the polarized neutron diffraction (PND) data at 10 K. Listed are the parameters for the moments in a chemical unit cell. Each component of their magnetic moments is given by $M_i = A_i \cos(\phi_i)$. The magnetic moments of the corresponding ions in a chemical unit cell separated by a distance $R$ from the original unit cell can be estimated by $M_i = A_i \cos(k_{IC} \cdot R + \phi_i)$. Errors for all parameters are given within parenthesis.

| | Position | $A_a$ ($\mu_B$) | $\phi_a$ (rad) | $A_b$ ($\mu_B$) | $\phi_b$ (rad) | $A_c$ ($\mu_B$) | $\phi_c$ (rad) | M ($\mu_B$) |
|---|---|---|---|---|---|---|---|---|
| Mn$^{3+}$ | (0.088, 0.851, 0.5) | 2.37(6) | 0.99(6) | 1.84(6) | -0.41(5) | 0.26(5) | 1.16(24) | 2.1(2) |
| Mn$^{3+}$ | (0.912, 0.149, 0.5) | 1.57(9) | 1.06(7) | 1.52(6) | -0.26(6) | 1.04(6) | 1.34(8) | 1.7(1) |
| Mn$^{3+}$ | (0.412, 0.351, 0.5) | -2.79(7) | -0.16(6) | -1.36(6) | -1.65(7) | 0.84(6) | -1.32(10) | 2.8(1) |
| Mn$^{3+}$ | (0.588, 0.649, 0.5) | 2.61(8) | -0.13(6) | 1.17(7) | -1.91(8) | 0.73(6) | 0.75(9) | 2.7(1) |
| Mn$^{4+}$ | (0.5, 0, 0.255) | -1.32(9) | 0.87(6) | -1.59(6) | -0.74(4) | -0.66(7) | 0.96(10) | 1.5(2) |
| Mn$^{4+}$ | (0, 0.5, 0.255) | 1.91(7) | -0.49(5) | 1.05(6) | -2.30(7) | 0.61(6) | 0.88(12) | 1.9(2) |
| Mn$^{4+}$ | (0, 0.5, 0.745) | 1.69(7) | 0.17(5) | 0.98(5) | -1.45(8) | -0.89(6) | -1.63(9) | 1.7(1) |
| Mn$^{4+}$ | (0.5, 0, 0.745) | -1.26(8) | 1.51(5) | -1.45(6) | 0.01(4) | -0.50(7) | -4.11(12) | 1.5(1) |

**Figure captions**

**Fig. 1.** FCD and PND data obtained from a single crystal of YMn$_2$O$_5$ (a), (b) at 25 K (ITC-FE phase) and (c), (d) at 10 K (LTI-FE). (a) and (c) show the unpolarized neutron scattering intensities of magnetic Bragg reflections obtained by our FCD measurements, while (b) and (d) show the data obtained from our CRYOPAD measurements. $I_{SF}$ and $I_{NSF}$ represent the spin-flip and the non-spin-flip scattering intensity, respectively, measured with neutrons polarized vertically to the (h0l) scattering plane. Black squares are the experimental data and are underneath red symbols that are calculated values for the complex spiral structures whose parameters are listed in Table I and II and illustrated in Fig. 2 and 3. Blue diamonds are calculated values for the FCD and PND data for the nearly collinear and coplanar magnetic structures reported by a previous neutron powder diffraction study. Cyan diamonds in (b) are the calculated PND data based on the weakly spiral commensurate magnetic structure that was reported by previous FCD studies. The goodness of fit, $\chi^2$, was defined by

$$\chi^2 = \frac{1}{N_p}\left(\sum_{FCD}\left(\frac{I_{obs}-I_{cal}}{\Delta I_{obs}}\right)^2 + \sum_{PND}\left(\frac{P^{zz}_{obs}-P^{zz}_{cal}}{\Delta P^{zz}_{obs}}\right)^2\right)$$

where $N_p$ is the number of the fitted parameters, $I_{obs}$ is the FCD intensity, and $P^{zz}_{obs} = \frac{I_{NSF}-I_{SF}}{I_{NSF}+I_{SF}}$ is the PND data. $I_{cal}$ and $P^{zz}_{cal}$ are the calculated values for $I_{obs}$ and $P^{zz}_{obs}$, respectively. The optimal $\chi^2$ was 24.6 and 8.3 for our commensurate (red symbols in (a),(b)) and incommensurate (red symbols in (c),(d)) spin structures, respectively.

**Fig. 2.** Magnetic structure of the ITC-FE phase of YMn$_2$O$_5$ that is projected onto different planes: (a) ab-plane, (b) ac-plane, (c) bc-plane. In (a)-(c), only Mn$^{4+}$ (red spheres) and surrounding oxygens (yellow spheres) are shown. In (d), Mn$^{4+}$ and Mn$^{3+}$ (green spheres) ions are shown in the ab-plane projection. The thin black rectangle in each figure represents a chemical unit cell. Blue curves represent spirals formed by the magnetic ions. Thick blue arrows represent electric polarizations induced by the spirals. The length of the arrow is scaled to the magnitude of **P**.

15**Fig. 3.** Magnetic structure of the LTI-FE phase of YMn$_2$O$_5$ that is projected onto different planes: (a) ab-plane, (b) ac-plane, (c) bc-plane, (d) ab-plane. In (d) thin grey lines represent the Mn$^{4+}$-Mn$^{3+}$-Mn$^{4+}$ zigzag chains. The symbols and colors were used in the same way as in Fig. 2.

Fig. 1

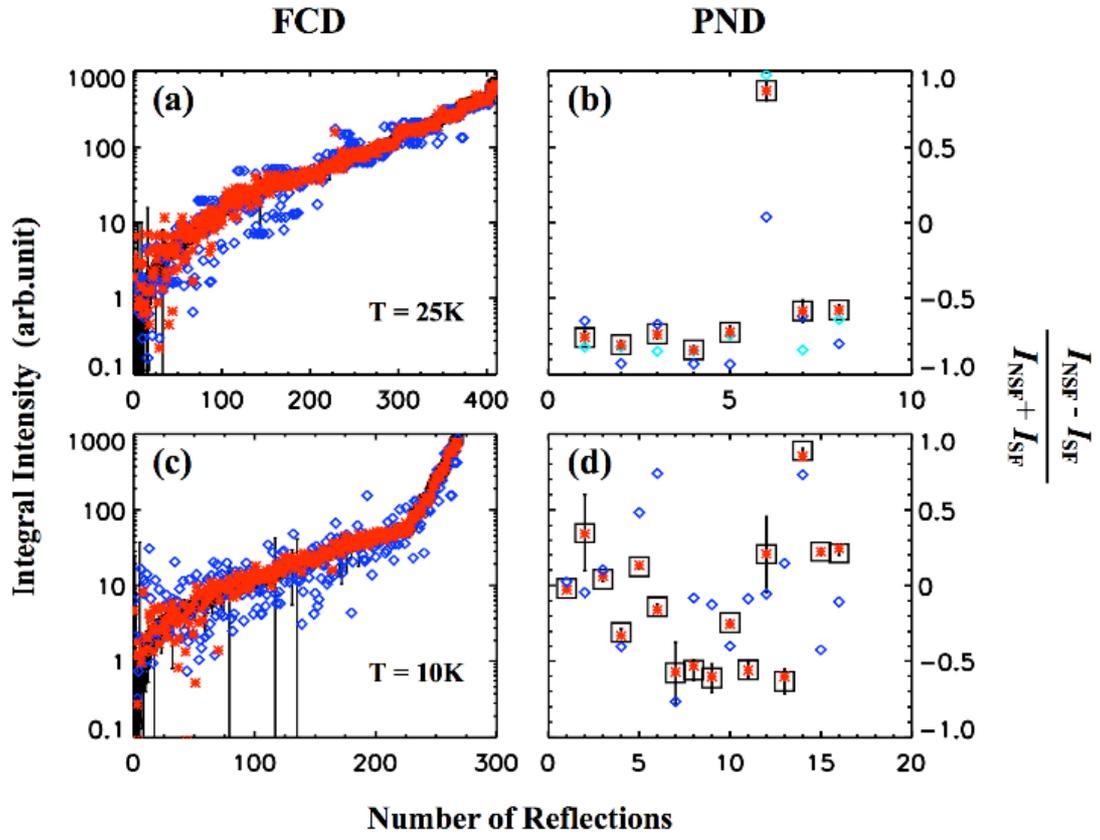

Fig. 2

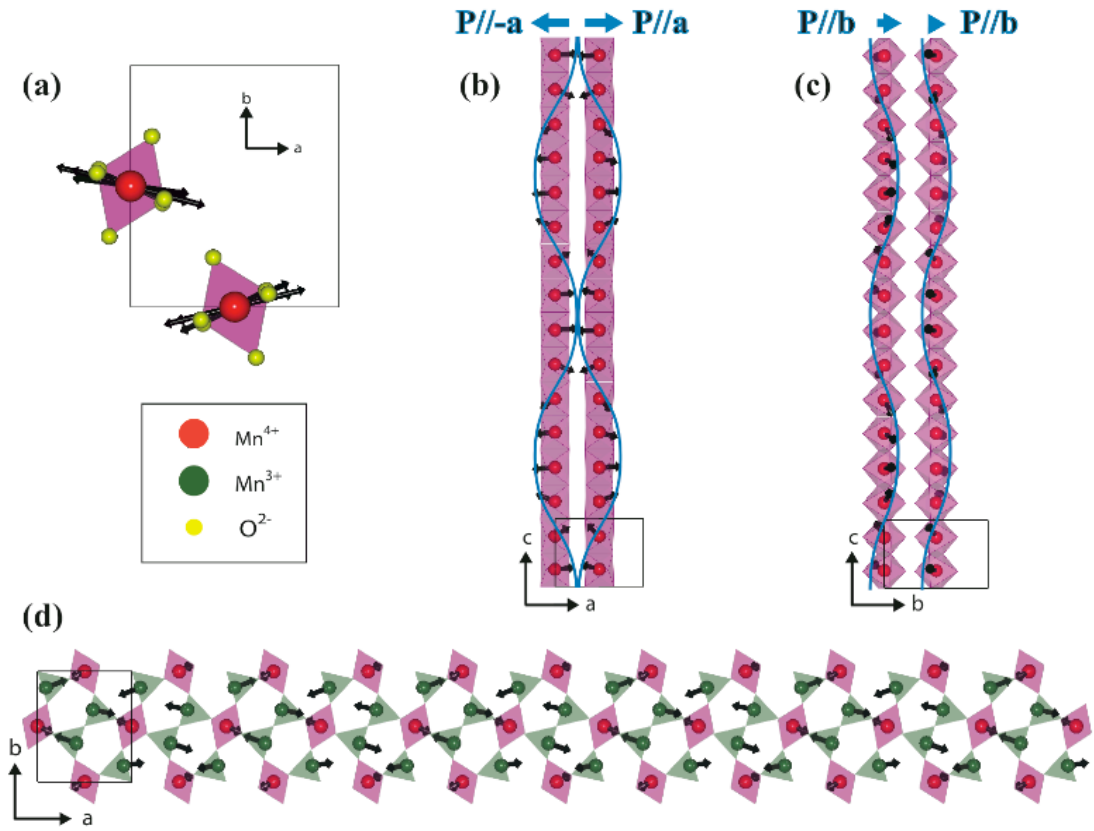

Fig. 3

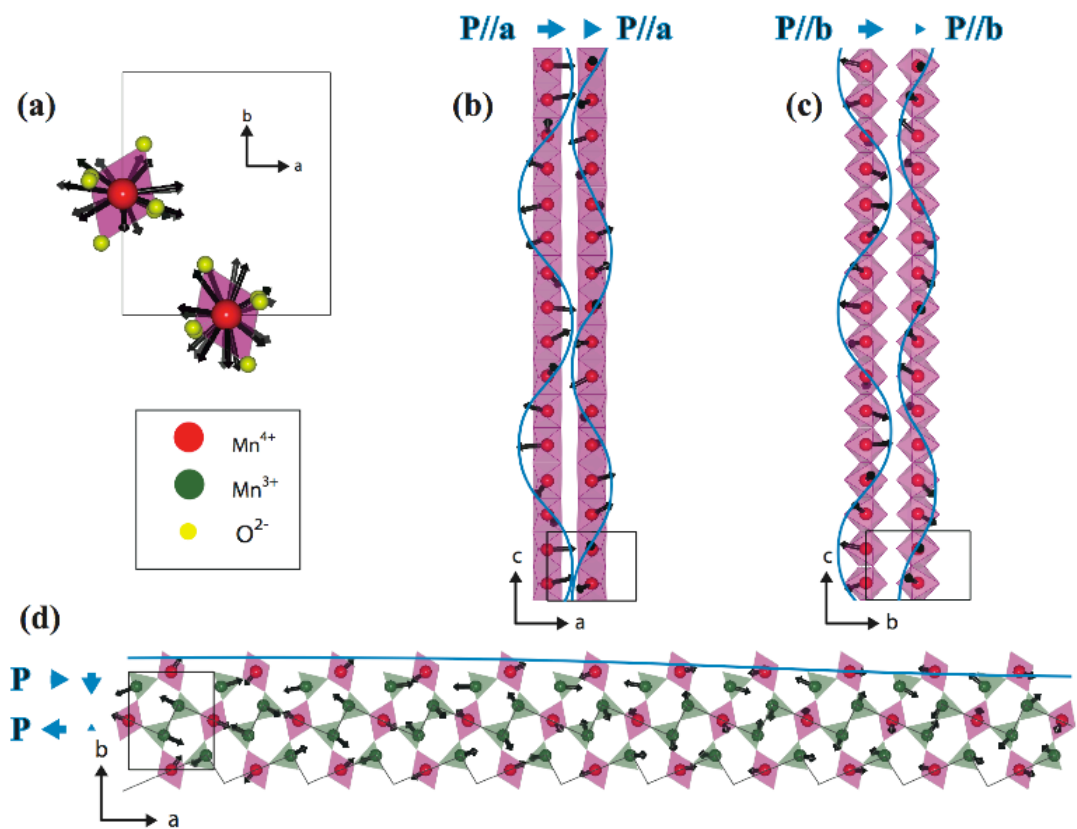